\documentclass[fleqn,usenatbib]{mnras}

\usepackage{newtxtext,newtxmath}

\usepackage[T1]{fontenc}

\DeclareRobustCommand{\VAN}[3]{#2}
\let\VANthebibliography\thebibliography
\def\thebibliography{\DeclareRobustCommand{\VAN}[3]{##3}\VANthebibliography}

\usepackage{graphicx}	
\usepackage{amsmath}	

\usepackage{amssymb}	

\usepackage{bm}
\usepackage{subfig}
\newcommand{\ksm}{~{\rm km}~{\rm s}^{-1}~{\rm Mpc}^{-1}}

\title[Constraints on IDE models from TD cosmography]{Constraints on interacting dark energy models from time-delay cosmography with seven lensed quasars}

\author[Ling-Feng Wang et al.]{
Ling-Feng Wang,$^{1}$
Jie-Hao Zhang,$^{1}$
Dong-Ze He,$^{2}$
Jing-Fei Zhang$^{1}$
and Xin Zhang$^{1,3,4}$\thanks{E-mail: zhangxin@mail.neu.edu.cn}
\\
$^{1}$Department of Physics, College of Sciences, Northeastern University, Shenyang 110819, China\\
$^{2}$College of Sciences, Chongqing University of Posts and Telecommunications, Chongqing 400065, China\\
$^{3}$Frontiers Science Center for Industrial Intelligence and Systems Optimization, Northeastern University, Shenyang 110819, China\\
$^{4}$Key Laboratory of Data Analytics and Optimization for Smart Industry (Northeastern University), Ministry of Education, Shenyang 110819, China
}

\date{Accepted 2022 May 24. Received 2022 May 21; in original form 2021 November 15}

\pubyear{2022}

\begin{document}
\label{firstpage}
\pagerange{\pageref{firstpage}--\pageref{lastpage}}
\maketitle

\begin{abstract}
Measurements of time-delay cosmography of lensed quasars can provide an independent probe to explore the expansion history of the late-time Universe. In this paper, we employ the time-delay cosmography measurements from seven lenses (here abbreviated as the TD data) to constrain interacting dark energy (IDE) models. We mainly focus on the scenario of vacuum energy (with $w=-1$) interacting with cold dark matter, and consider four typical cases of the interaction form $Q$. When the TD data alone are employed, we find that the IDE models with $Q\propto \rho_{\rm de}$ seem to have an advantage in relieving the $H_{0}$ tension between the cosmic microwave background (CMB) and TD data. When the TD data are added to the CMB$+$BAO$+$SN$+H_0$ data, we find that: (i) the coupling parameter $\beta$ in all the considered IDE models is positive within 1$\sigma$ range, implying a mild preference for the case of cold dark matter decaying into dark energy; (ii) the IDE model with $Q = \beta H_{0} \rho_{\rm c}$ slightly relieves the $S_8$ tension, but the other considered IDE models further aggravate this tension; (iii) the Akaike information criteria of the IDE models with $Q \propto \rho_{\rm c}$ are lower than that of the $\Lambda$CDM model, indicating that these IDE models are more preferred by the current mainstream data. We conclude that the considered IDE models have their own different advantages when the TD data are employed, and none of them can achieve good scores in all aspects.
\end{abstract}




\section{Introduction}

The precise measurement of the cosmic microwave background (CMB) anisotropies indicates the dawn of an era of precision cosmology \citep{Bennett:2003bz,Spergel:2003cb}. Six basic parameters in the standard $\Lambda$CDM model have been constrained precisely by the \emph{Planck} CMB observation \citep{Aghanim:2018eyx}, and the $\Lambda$CDM model can excellently fit most of the observational data with the least free parameters. Nevertheless, theoretically, the $\Lambda$CDM model suffers from the fine-tuning and cosmic coincidence problems \citep{Weinberg:1988cp}. {Furthermore, the Hubble constant $H_0$ inferred from the \emph{Planck} CMB observation combined with the $\Lambda$CDM model \citep{Aghanim:2018eyx} is above 5$\sigma$ tension with the direct measurements by the SH0ES (Supernovae H0 for the Equation of State) team \citep{Riess:2021jrx} using the distance ladder method (see reviews by e.g. \citealp{Knox:2019rjx,DiValentino:2021izs,DiValentino:2020zio,Jedamzik:2020zmd,Perivolaropoulos:2021jda,Shah:2021onj,Abdalla:2022yfr}).} Thus, some extensions to the $\Lambda$CDM model are expected to be necessary not only for the development of cosmological theories, but also for the interpretation of experimental results \citep{DiValentino:2020zio,Abdalla:2022yfr}.


Among various extensions to the $\Lambda$CDM model, a variety of models based on the scenario of dark energy interacting with cold dark matter, referred to as the interacting dark energy (IDE) models, have attracted lots of attention (see \citealp{Wang:2016lxa} for a recent review). This scenario could help resolve the cosmic coincidence problem \citep[e.g.][]{Comelli:2003cv,Cai:2004dk,Zhang:2005rg,He:2008tn,He:2009pd} and relieve the $H_0$ tension \citep[e.g.][]{DiValentino:2017iww,Yang:2018euj,Yang:2018uae,Pan:2019gop,DiValentino:2019ffd,DiValentino:2019jae,Vagnozzi:2019ezj,Gao:2021xnk}.
In addition, the knowledge of dark energy and dark matter is still very scarce, and whether there exists a direct, non-gravitational interaction between them has yet to be verified. Therefore, indirectly detecting the interaction between dark sectors via cosmological observations is one of the important missions in current cosmology \citep[e.g.][]{Guo:2007zk,Li:2009zs,Xia:2009zzb,He:2010im,Li:2010ak,Li:2011ga,Fu:2011ab,Zhang:2012uu,Zhang:2013lea,Cui:2015ueu,Feng:2016djj,Murgia:2016ccp,Xia:2016vnp,Costa:2016tpb,Guo:2018gyo,Li:2018ydj,Asghari:2020ffe,Cheng:2019bkh,Feng:2019jqa,Li:2020gtk,Pan:2020mst,Aljaf:2020eqh,Carrilho:2021hly,DiValentino:2020kpf,Lucca:2021eqy,Samart:2021viu,Yang:2021hxg}.
Describing the interaction between dark sectors usually requires the introduction of extra parameters in the base $\Lambda$CDM model, which cannot be precisely constrained by the CMB data, because these extra parameters are highly degenerate with other cosmological parameters. For the sake of breaking the cosmological parameter degeneracies and constraining the interaction between dark sectors more precisely, a variety of precise late-Universe probes are required \citep[e.g.][]{Yang:2019vni,Li:2019ajo,Zhang:2021yof}.

Time-delay cosmography is a promising late-Universe probe, providing independent measurements for $H_0$ by measuring the time delays between the multiple images of gravitational lenses \citep{Treu:2016ljm}. Compared with the distance ladder, this method measures the combination of absolute angular diameter distances rather than relative distances, so it can provide constraints on $H_0$.
In 1964, \citet{Refsdal:1964nw} proposed that the angular diameter distance of the lens system could be inferred by measuring the arrival time difference $\Delta t$ for the light rays from different paths. Active galactic nucleus is the background source that can provide a sufficiently variable luminosity to measure the time delay \citep{Vanderriest:1989aa,Schechter:1996fa,Fassnacht:1999re,Kochanek:2005ge,Eigenbrod:2005yy}. In 1979, the first strongly lensed quasar with two images was discovered \citep{Walsh:1979nx}, and the first robust time delays were measured in 1997 \citep{Kundic:1996tr}. Whereafter, time-delay cosmography gradually becomes a high-profile cosmological probe aimed at the late-time Universe \citep[e.g.][]{Fassnacht:2002df,Kochanek:2002rk,Koopmans:2003ha,Eigenbrod:2005ub,Fassnacht:2005uc,Paraficz:2009xj,Suyu:2012aa,Jee:2014uxa,Meng:2015qia,Bonvin:2015jia,Chen:2016fwu,Collett:2016muz,Jee:2015yra}.

Recently, the H0LiCOW (H0 Lenses in COSMOGRAIL’s Wellspring) collaboration employed the time-delay cosmography measurements with six lensed quasars to infer the $H_0$ value \citep{Wong:2019kwg}. The inferred $H_{0}$ value in a spatially flat $\Lambda$CDM cosmology is $73.3^{+1.7}_{-1.8} \ksm$, which is in $3.1 \sigma$ tension with the $\emph{Planck}$ CMB result. This tension further increases to 5.3$\sigma$ when combining the time-delay cosmography with the local $H_0$ measurement by the SH0ES team.
\citet{Wong:2019kwg} have considered several extended cosmological models to relieve the $H_0$ tension, and found that the tension still exists in those extended models.
Then, the STRIDES (STRong-lensing Insights into Dark 
Energy Survey) collaboration employed the time-delay cosmography from the single lens DES\,J0408$-$5354 to infer the $H_0$ value, and obtained $H_0=74.2^{+2.7}_{-3.0} \ksm$ with precision of 3.9\% \citep{DES:2019fny}, which is in 2.2$\sigma$ tension with the $\emph{Planck}$ CMB result.
Whereafter, the TDCOSMO (Time-Delay COSMOgraphy) collaboration reanalysed the lensed quasars in the H0LiCOW and STRIDES samples to investigate the effect of lens mass models on the systematic uncertainties \citep{Birrer:2020tax,Millon:2019slk}, and obtained $H_0=74.0^{+1.7}_{-1.8} \ksm$ (composite model) and $H_0=74.2\pm1.6 \ksm$ (power-law model), which are consistent with the previous results. The $H_0$ tension between the $\emph{Planck}$ CMB observation and the time-delay cosmography reflects the inconsistency of the measurements between the early and late Universe.
The IDE models may have the potential to relieve the $H_0$ tension by introducing the interaction between dark sectors, but a detailed analysis on various IDE models with the time-delay cosmography is lacking so far.

In the light of the introduction above, we find that there are several important questions that deserve to be studied: (i) which type of the IDE models has advantage in relieving the $H_0$ tension between the time-delay cosmography and the CMB observation; (ii) how the time-delay cosmography affects the cosmological constraints on the interaction between dark energy and cold dark matter; (iii) which type of the IDE models is more favoured by the current mainstream data sets including the time-delay cosmography. To answer these questions, in this paper, we employ the time-delay cosmography measurements from seven lensed quasars to constrain several typical IDE models.

The rest of this paper is organized as follows. In Section~\ref{sec:Method}, we introduce the theoretical basis of the time$-$delay cosmography and the IDE models, and show the observational data considered in this work. In Section~\ref{sec:Result}, we report the constraint results and give detailed discussions. Finally, the conclusion is given in Section~\ref{sec:con}.


\begin{table*}
    \renewcommand\arraystretch{1.5}
	\centering
	\caption{The TD data from seven lensed quasars, including the measurements of $z_{\rm d}$, $z_{\rm s}$, and $D_{\Delta t}$. Here, $z_{\rm s}$ is the redshift of the source.}
	\label{7lens}
	\begin{tabular}{lccr} 
		\hline
        Lens name & $z_{\rm d}$ & $z_{\rm s}$ & $D_{\Delta t}$(Mpc)\\
        \hline
        B1608$+$656 \citep{Suyu:2009by,Jee:2019hah} & $0.6304$ & $1.394$ & $5156^{+296}_{-236}$\\
        RXJ1131$-$1231 \citep{Suyu:2013kha,Chen:2019ejq} & $0.295$ & $0.654$ & $2096^{+98}_{-83}$\\
        HE\,0435$-$1223 \citep{Wong:2016dpo,Chen:2019ejq} & $0.4546$ & $1.693$& $2707^{+183}_{-168}$\\
        SDSS\,1206$+$4332 \citep{Birrer:2018vtm} & $0.745$ & $1.789$ & $5769^{+589}_{-471}$\\
        WFI2033$-$4723 \citep{Rusu:2019xrq} & $0.6575$ & $1.662$ & $4784^{+399}_{-248}$\\
        PG\,1115$+$080 \citep{Chen:2019ejq} & $0.311$ & $1.722$& $1470^{+137}_{-127}$\\
        DES\,J0408$-$5354 \citep{Agnello:2017mwu,DES:2019fny} & $0.597$ & $2.375$ & $3382^{+146}_{-115}$\\
        \hline
	\end{tabular}
\end{table*}

\section{METHODS AND DATA}\label{sec:Method}

\subsection{Time-delay cosmography}\label{Time-Delay}
When a massive object (the lens) lies between a background source and an observer, the background source may be gravitationally lensed into multiple images. The light rays corresponding to different image positions travel through different space-time paths. Since these paths have different gravitational potentials and lengths, these light rays will reach the observer at different times. If the source has flux variations, the time delays between multiple images can be measured by monitoring the lens \citep{Schechter:1996fa,Fassnacht:1999re,Fassnacht:2002df,Kochanek:2005ge,Courbin:2010au}.

If the foreground lens and the background source are sufficiently aligned, multiple images of the background source may be formed. The light rays observed at different images have different excess time delays when they reach the observer.
The excess time delay between two images is defined by
\begin{equation} \label{eq:td}
\Delta t_{ij} = \frac{D_{\Delta t}}{c} \left[ \frac{(\bm{\theta}_{i} - \bm{\beta})^{2}}{2} - \psi(\bm{\theta}_{i}) - \frac{(\bm{\theta}_{j} - \bm{\beta})^{2}}{2} + \psi(\bm{\theta}_{j}) \right],
\end{equation}
where $\bm{\theta}_{i}$ and $\bm{\theta}_{j}$ are the positions of images $i$ and $j$ in the image plane, respectively. The lens potentials at the image positions, $\psi(\bm{\theta}_{i})$ and $\psi(\bm{\theta}_{j})$, and the source position $\bm{\beta}$, can be determined from the mass model of the system. The time-delay distance $D_{\Delta t}$ \citep{Refsdal:1964nw,Suyu:2009by} is defined as a combination of three angular diameter distances,
\begin{equation} \label{eq:ddt}
D_{\Delta t} \equiv (1+z_{\rm d}) \frac{D_{\rm d} D_{\rm s}}{D_{\rm ds}},
\end{equation}
where $z_{\rm d}$ is the redshift of the lens, $D_{\rm d}$ is the angular diameter distance to the lens, $D_{\rm s}$ is the angular diameter distance to the source, and $D_{\rm ds}$ is the angular diameter distance between the lens and the source. If the time delay $\Delta t_{ij}$ can be measured and an accurate lens model is available to determine the lens potential $\psi(\bm{\theta})$, then the time-delay distance can be determined. By further assuming a cosmological model, $D_{\Delta t}$ can be used to constrain cosmological parameters.

\subsection{Interacting dark energy models}\label{models}
In the context of a spatially flat Friedmann$-$Roberston$-$Walker universe, the Friedmann equation is written as
\begin{equation}\label{2.1}
3M^2_{\rm{pl}} H^2=\rho_{\rm{de}}+\rho_{\rm c}+\rho_{\rm b}+\rho_{\rm r},
\end{equation}
where $3M^2_{\rm{pl}} H^2$ is the critical density of the Universe, $\rho_{\rm{de}}$, $\rho_{\rm c}$, $\rho_{\rm b}$, and $\rho_{\rm r}$ represent the energy densities of dark energy, cold dark matter, baryon, and radiation, respectively.

In the IDE models, the assumption of some direct, non-gravitational interaction between dark energy and cold dark matter is made. Under this assumption, in the level of phenomenological study, the energy conservation equations for dark energy and cold dark matter are given by
\begin{align}\label{conservation1}
\dot{\rho}_{\rm de} +3H(1+w)\rho_{\rm de}= Q,\\
\dot{\rho}_{\rm c} +3 H \rho_{\rm c}=-Q,
\end{align}
where the dot denotes the derivative with respect to the cosmic time $t$, $w$ is the equation of state parameter of dark energy, and $Q$ describes the energy transfer rate between dark sectors.

Since the fundamental nature of dark energy is still unclear, it is difficult to understand the microscopic origin of the interaction between dark energy and cold dark matter. Therefore, we can only study the IDE in a pure phenomenological way \citep[e.g.][]{Zhang:2005rg,Zhang:2007uh,Zhang:2009qa,Li:2010ak,Li:2013bya,Zhang:2013lea,Li:2014eha,Li:2014cee,Geng:2015ara,Li:2015vla}. The form of $Q$ is usually assumed to be proportional to the energy density of dark energy or cold dark matter, or some mixture of the two \citep[e.g.][]{Amendola:1999qq,Billyard:2000bh}. The proportionality coefficient $\Gamma$ has the dimension of energy, and so it is with the form of $\Gamma=\beta H$ or $\Gamma=\beta H_0$, where $\beta$ is the dimensionless coupling parameter. $\beta = 0$ indicates no interaction between dark energy and cold dark matter, $\beta > 0$ means cold dark matter decaying into dark energy, and $\beta < 0$ means dark energy decaying into cold dark matter.

In this work, we study the minimal version of extension to the base $\Lambda$CDM model in the context of IDE. Hence, we consider only the case of $w=-1$, in order not to introduce more extra parameters. If there is no interaction between dark sectors, the case of $w=-1$ corresponds to the vacuum energy, serving as a pure background in the cosmological evolution. However, when there is some interaction between dark sectors, even though we have $w=-1$, the corresponding dark energy cannot serve as a pure background and it actually is not a vacuum energy in essence. Here we do not wish to study the nature of dark energy with a purely theoretical point of view, but instead we wish to study the problem concerning dark energy in a phenomenological way. Thus, we call the dark energy with $w=-1$ vacuum energy for convenience in this paper, and the corresponding IDE model is denoted as the I$\Lambda$CDM model. In this work, we take four specific forms of $Q$ as typical examples to make an analysis of the IDE models, i.e., $Q=\beta H_{0}\rho_{\rm c}$ (I$\Lambda$CDM1), $Q=\beta H_{0}\rho_{\rm de}$ (I$\Lambda$CDM2), $Q=\beta H\rho_{\rm c}$ (I$\Lambda$CDM3), and $Q=\beta H\rho_{\rm de}$ (I$\Lambda$CDM4).

In the IDE cosmology, the early-time superhorizon cosmological perturbations occasionally diverge (in a part of the parameter space of the model) if the dark energy perturbations are considered, leading to a catastrophe in cosmology as the perturbations enter the horizon. In order to avoid such a cosmological catastrophe caused by the interaction between dark sectors, one has to consider some effective schemes to properly calculate the perturbations of dark energy, instead of using a conventional way of treating dark energy as a perfect fluid with negative pressure.
In 2014, \citet{Li:2014eha,Li:2014cee} extended the parametrized post-Friedmann (PPF) approach \citep{Fang:2008sn,Hu:2008zd} to involve the IDE. Such an extended version of PPF method, referred to as ePPF for convenience, can successfully avoid the perturbation divergence problem in the IDE cosmology. In this work, we consider the I$\Lambda$CDM model in which the ``vacuum energy'' is not a true background due to the interaction, thus we also need to consider its perturbations. Therefore, we employ the ePPF method \citep{Li:2014eha,Li:2014cee} to treat the cosmological perturbations in this work (see e.g. \citealp{Li:2015vla,Zhang:2017ize,Feng:2018yew} for more applications of the ePPF method).

\begin{table*}
    \renewcommand\arraystretch{1.5}
	\centering
	\caption{Fitting results (68.3 per cent confidence level) in the $\Lambda$CDM and IDE models from the CMB and TD data. Here, $H_{0}$ is in units of ${\rm km}~{\rm s}^{-1}~{\rm Mpc}^{-1}$.}
	\label{TD}
	\begin{tabular}{lcccccr} 
	    \hline
    	Data&Parameter & $\Lambda$CDM & I$\Lambda$CDM1 & I$\Lambda$CDM2 & I$\Lambda$CDM3 & I$\Lambda$CDM4\\
		\hline
		CMB & $H_{0}$ & $67.15\pm 0.61$ & $60.4^{+4.5}_{-5.3}$ & $67.3\pm 3.2$ & $65.5\pm 1.8$ & $66.1\pm 4.6$\\
		& $\Omega _{m}$ & $0.3177\pm 0.0085$ & $0.51^{+0.11}_{-0.19}$ & $ 0.32\pm 0.13$ & $0.341\pm 0.025$ & $0.35\pm 0.15$\\
		& $\beta$ & $-$ & $-0.31\pm 0.21$ & $-0.04^{+0.70}_{-0.44}$ & $-0.0023\pm 0.0023$ & $-0.11^{+0.54}_{-0.44}$\\
		TD & $H_{0}$ & $73.8\pm 1.6$ & $ 74.3\pm 1.9 $ & $74.3\pm 1.8$ & $ 74.1\pm 1.7$ & $ 74.2\pm 1.8$\\
		& $\Omega _{m}$ & $0.249^{+0.022}_{-0.028}$ & $0.258^{+0.025}_{-0.017}$ & $0.253^{+0.024}_{-0.018}$ & $0.263^{+0.013}_{-0.019}$ & $0.261\pm 0.016$\\
		& $\beta$ & $-$ & $ > -0.246$ & $ 0.21^{+0.50}_{-0.46}$ & $ -0.11^{+0.50}_{-0.36}$ & $ 0.07\pm 0.31$\\
		\hline
		& $H_{0} $ tension & $3.88\sigma$ & $2.64\sigma$ & $1.91\sigma$ & $3.47\sigma$ & $1.64\sigma$\\
        \hline
	\end{tabular}
\end{table*}

\begin{figure*}
	\subfloat[]{\label{fig1a}\includegraphics[width=0.9\columnwidth]{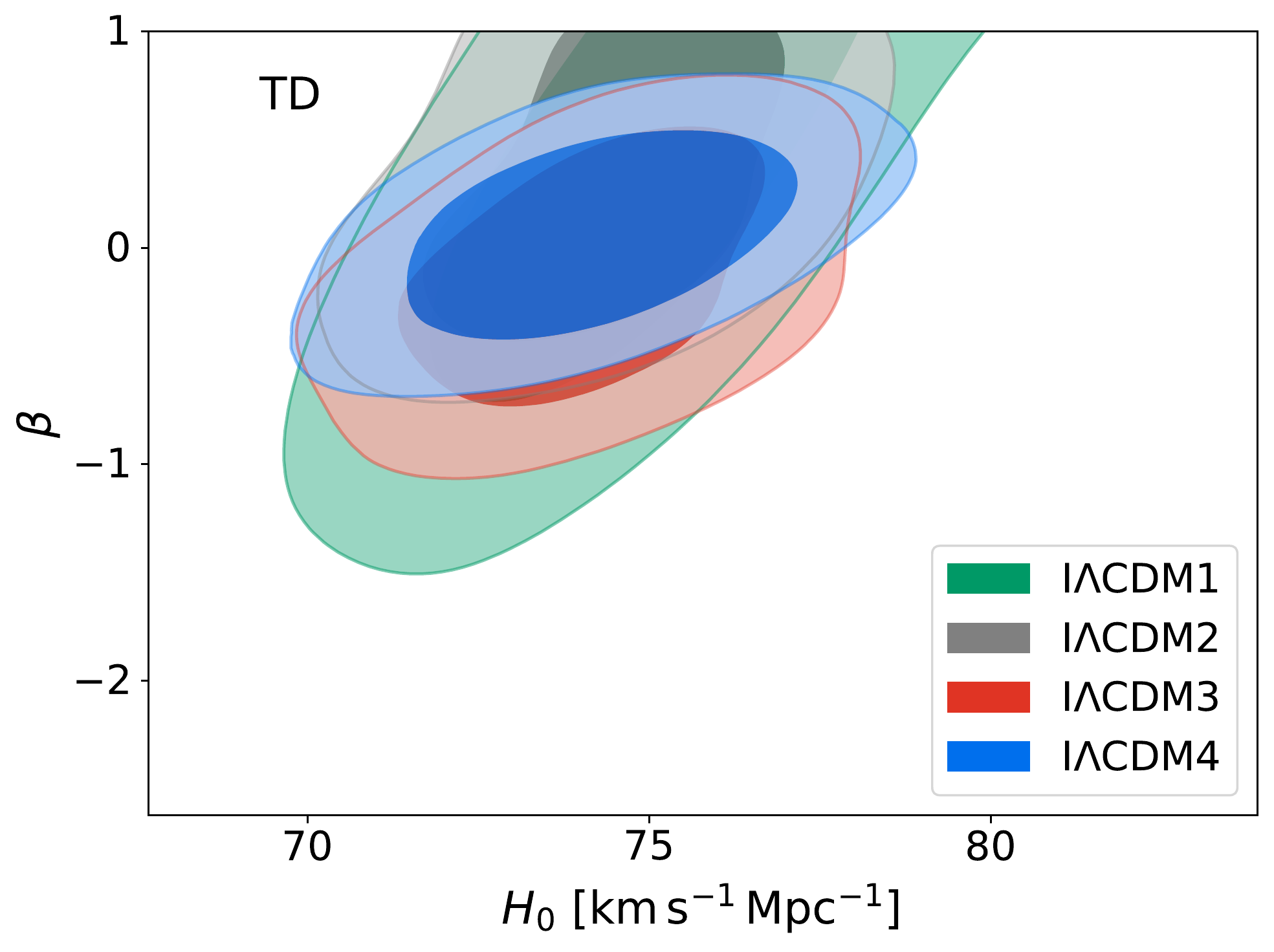}}
    \subfloat[]{\label{fig1b}\includegraphics[width=0.9\columnwidth]{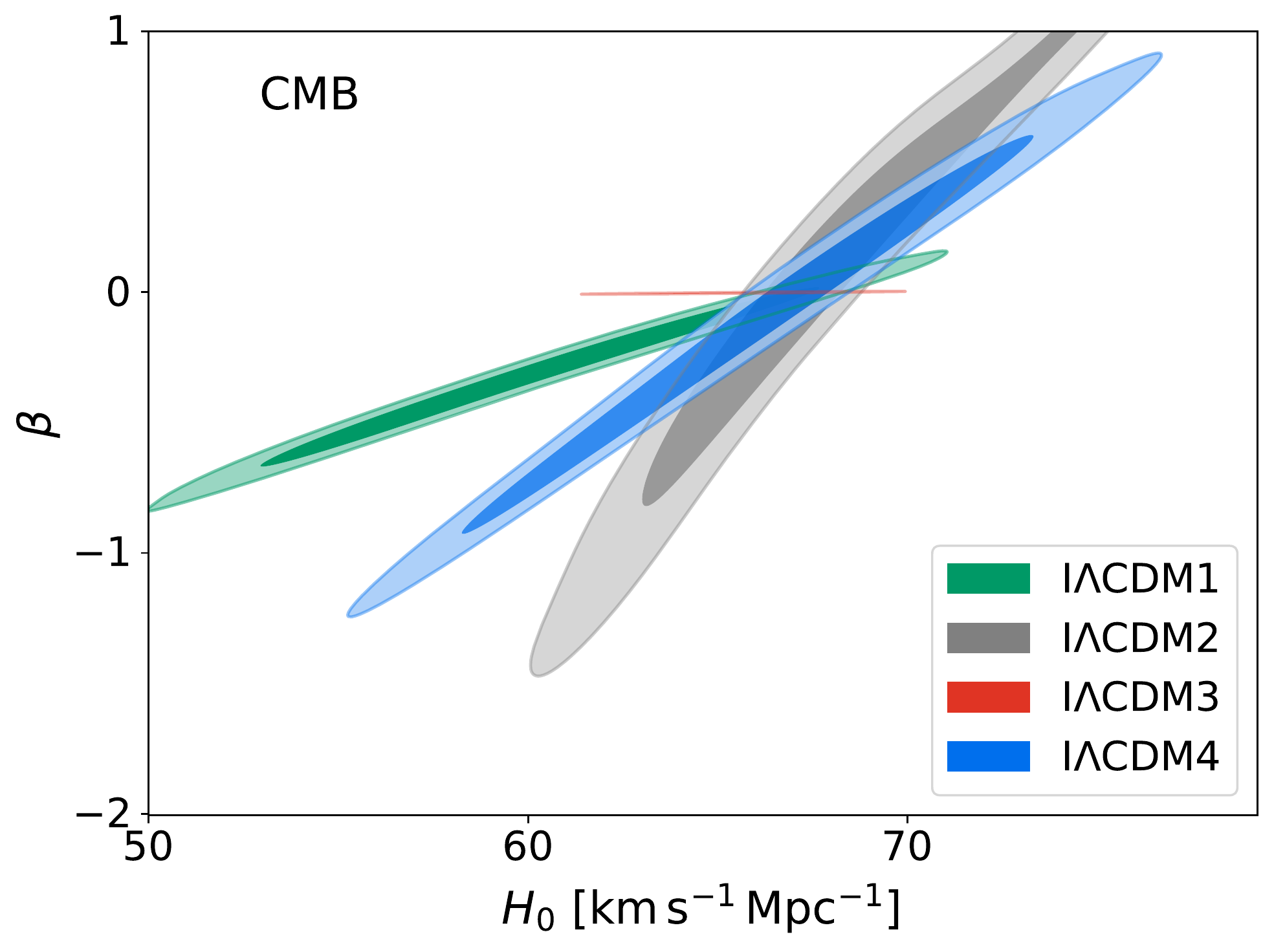}}
    \caption{2D marginalized contours (68.3 and 95.4 per cent confidence levels) in the $H_{0}$--$\beta$ planes for the IDE models by using the TD and CMB data.} 
    \label{fig1}
\end{figure*}

\subsection{Observational data}\label{data}
We employ the modified version of the Markov Chain Monte Carlo package {\tt COSMOMC} \citep{Lewis:2002ah} to infer the posterior distributions of the cosmological parameters. The observational data used in this paper include the CMB data, the baryon acoustic oscillation (BAO) data, the type Ia supernova (SN) data, the $H_0$ data, the galaxy clustering (GC) and weak lensing (WL) data, and the time-delay cosmography data. Unless otherwise specified, we use the abbreviation `TD' to represent the time-delay cosmography data in the following. The details of these data are listed as follows.

\begin{itemize}
\item[(i)] The CMB data: the \emph{Planck} TT, TE, EE spectra at $\ell \geq 30$, the low-$\ell$ temperature Commander likelihood, and the low-$\ell$ SimAll EE likelihood, from the \emph{Planck} 2018 data release \citep{Aghanim:2018eyx}.

\item[(ii)] The BAO data: the measurements from 6dFGS ($z_{\rm eff}=0.106$) \citep{Beutler:2011hx}, SDSS-MGS ($z_{\rm eff}=0.15$) \citep{Ross:2014qpa}, and BOSS DR12 ($z_{\rm eff}=0.38$, 0.51, and 0.61) \citep{Alam:2016hwk}.

\item[(iii)] The SN data: the latest Pantheon sample comprised of 1048 data points from the Pantheon compilation \citep{Scolnic:2017caz}.

\item[(iv)] The $H_0$ data: the measurement result of $H_0=74.03\pm1.42$ $\ksm$ from distance ladder reported by the SH0ES team \citep{Riess:2019cxk}.

\item[(v)] The GC and WL data: the GC and WL data from the first year observation of the Dark Energy Survey  \citep[DES;][]{DES:2017myr}.

\item[(vi)] The TD data: the measurements of the time-delay cosmography from seven lensed quasars (six H0LiCOW lenses and one STRIDES lens), as listed in Table~\ref{7lens}.
\end{itemize}

\section{Results and discussions} \label{sec:Result}
In this section, we report our constraint results in detail, and make some analyses and discussions on them. First, we shall report the constraint results from the TD data alone, and make a comparison for the four types of IDE models in relieving the $H_0$ tension. Then, we add the local $H_0$ measurement by the SH0ES team to the TD data, for the sake of further showing the $H_0$ tension between the CMB and TD$+H_0$ data. Thirdly, we employ the CMB$+$BAO$+$SN$+H_0$ and CMB$+$BAO$+$SN$+H_0+$TD data to constrain the IDE models, in order to investigate the effect of adding the TD data to the mainstream observational data on the cosmological parameters. In addition, we shall also use the information criterion as a statistical tool to judge how well the IDE models can fit the data. We use the abbreviation `CBSH' to denote the CMB$+$BAO$+$SN$+H_0$ data.

We first report the constraint results from the TD data alone, which are shown in Table~\ref{TD}. For the parameter $H_0$, the TD data can constrain it with precision of $\sim2$ per cent in the $\Lambda$CDM and IDE models. The inferred $H_0$ value is $73.8\pm 1.6$ $\ksm$ in the $\Lambda$CDM model, which is in 3.88$\sigma$ tension with the CMB data. In the four IDE models, the inferred $H_0$ values are $74.3\pm 1.9$ (I$\Lambda$CDM1), $74.3\pm 1.8$ (I$\Lambda$CDM2), $74.1\pm 1.7$ (I$\Lambda$CDM3), and $74.2\pm 1.8$ (I$\Lambda$CDM4) $\ksm$. The $H_0$ tensions are relieved to a certain extent in all the IDE models, and especially in the I$\Lambda$CDM2 and I$\Lambda$CDM4 models, the tensions are reduced to 1.91$\sigma$ and 1.64$\sigma$, respectively.
However, these results actually do not support that the I$\Lambda$CDM2 and I$\Lambda$CDM4 models can relieve the $H_0$ tension.
In fact, the contours in Fig.~\ref{fig1}\subref{fig1a} show that there is no significant correlation between the parameters $\beta$ and $H_0$ when the TD data are employed, indicating that these IDE models cannot effectively change the $H_0$ values through introducing an interaction between the dark sectors. From Fig.~\ref{fig1}\subref{fig1b} and Table~\ref{TD} we can see that the reason why the IDE models seem to relieve the $H_0$ tension is the large errors of $H_0$ given by the CMB data, but not the better overlaps of parameter spaces. We note that in the IDE models, the constraints on the parameters are so loose that the posterior distributions obviously fluctuate with repeated analyses. Therefore, we choose the median values of multiple analyses as the final displayed results, and we verify that such an approach would not affect our main conclusions. {We also note that when only the CMB data are employed, the constraint results of $H_0$ in this paper are different from several previous papers \citep{DiValentino:2019ffd,DiValentino:2020vnx,Lucca:2020zjb}. For example, given the same IDE model (the I$\Lambda$CDM model with $Q\propto H\rho_{\rm de}$), the constraint results of $H_0$ are $66.1\pm 4.6$ $\ksm$ in this paper but $72.8^{+3.0}_{-1.5}$ $\ksm$ in \citet{DiValentino:2019ffd}. This difference is arising from the different treatments for cosmological perturbations. To avoid the perturbation divergence problem existing in the IDE cosmology, the priors of $\beta$ and $w$ are set to $\beta>0$ and $w>-1$ (in this case $w$ is fixed to $w=-0.999$) in \citet{DiValentino:2019ffd}, but the ePPF method used in this paper allows us to explore the whole parameter space without assuming any specific priors on $w$ and $\beta$.}

\begin{table*}
    \renewcommand\arraystretch{1.5}
	\centering
	\caption{Fitting results (68.3 per cent confidence level) in the $\Lambda$CDM and IDE models from the TD$+H_0$ data. Here, $H_{0}$ is in units of ${\rm km}~{\rm s}^{-1}~{\rm Mpc}^{-1}$.}
	\label{TD+H0}
	\begin{tabular}{lccccr} 
	    \hline
	    Parameter & $\Lambda$CDM & I$\Lambda$CDM1 & I$\Lambda$CDM2 & I$\Lambda$CDM3 & I$\Lambda$CDM4\\
	    \hline
	    $H_{0}$ & $ 73.9\pm 1.1$ & $74.2\pm 1.2$ & $74.2\pm 1.0$ & $74.1\pm 1.0$ & $74.0\pm 1.1$\\
	    $\Omega _{m}$ & $0.224^{+0.044}_{-0.033}$ & $0.262\pm 0.020$ & $0.260^{+0.013}_{-0.016}$ & $0.2603^{+0.0088}_{-0.015}$ & $0.261^{+0.015}_{-0.017}$\\
	    $\beta$ & $-$ & $> -0.162$ & $0.24^{+0.37}_{-0.28}$ & $0.06^{+0.27}_{-0.22}$ & $0.10^{+0.23}_{-0.30}$\\
	    \hline
	    $H_{0}$ tension & $5.37\sigma$ & $2.74\sigma$ & $2.06\sigma$ & $4.18\sigma $ & $1.67\sigma$\\
        \hline
	\end{tabular}
\end{table*}

Then, for the sake of further showing the $H_0$ tension between the CMB observation and the late-Universe observations, we combine the TD data with the local $H_0$ measurement by the SH0ES team to give constraints on the IDE models. The constraint results from the TD$+H_0$ data are listed in Table~\ref{TD+H0}. 
Because adding the local $H_0$ measurement in the data set is just equivalent to putting a prior on $H_0$, the constraint errors of $H_0$ become smaller, but the constraint errors of $\beta$ are still too large. There is still no significant correlation between the parameters $\beta$ and $H_0$, similar to the case of using the TD data alone. Specifically, the $H_0$ value is $73.9\pm1.1$ $\ksm$ in the $\Lambda$CDM model, which is in 5.37$\sigma$ tension with the value inferred from the CMB data. In the four IDE models, the $H_0$ value are $74.2\pm1.2$ (I$\Lambda$CDM1), $74.2\pm1.0$ (I$\Lambda$CDM2), $74.1\pm1.0$ (I$\Lambda$CDM3), and $74.0\pm1.1$ (I$\Lambda$CDM4) $\ksm$. Compared with the 5.37$\sigma$ tension in the $\Lambda$CDM model, the $H_0$ tensions are reduced to 2.06$\sigma$ and 1.67$\sigma$ in the I$\Lambda$CDM2 and I$\Lambda$CDM4 model, respectively. As discussed above, these reductions of $H_0$ tensions are actually due to the large errors of $H_0$ from the CMB data, but not the better overlaps of parameter spaces.

\begin{figure}
    \centering
	\includegraphics[width=0.9\columnwidth]{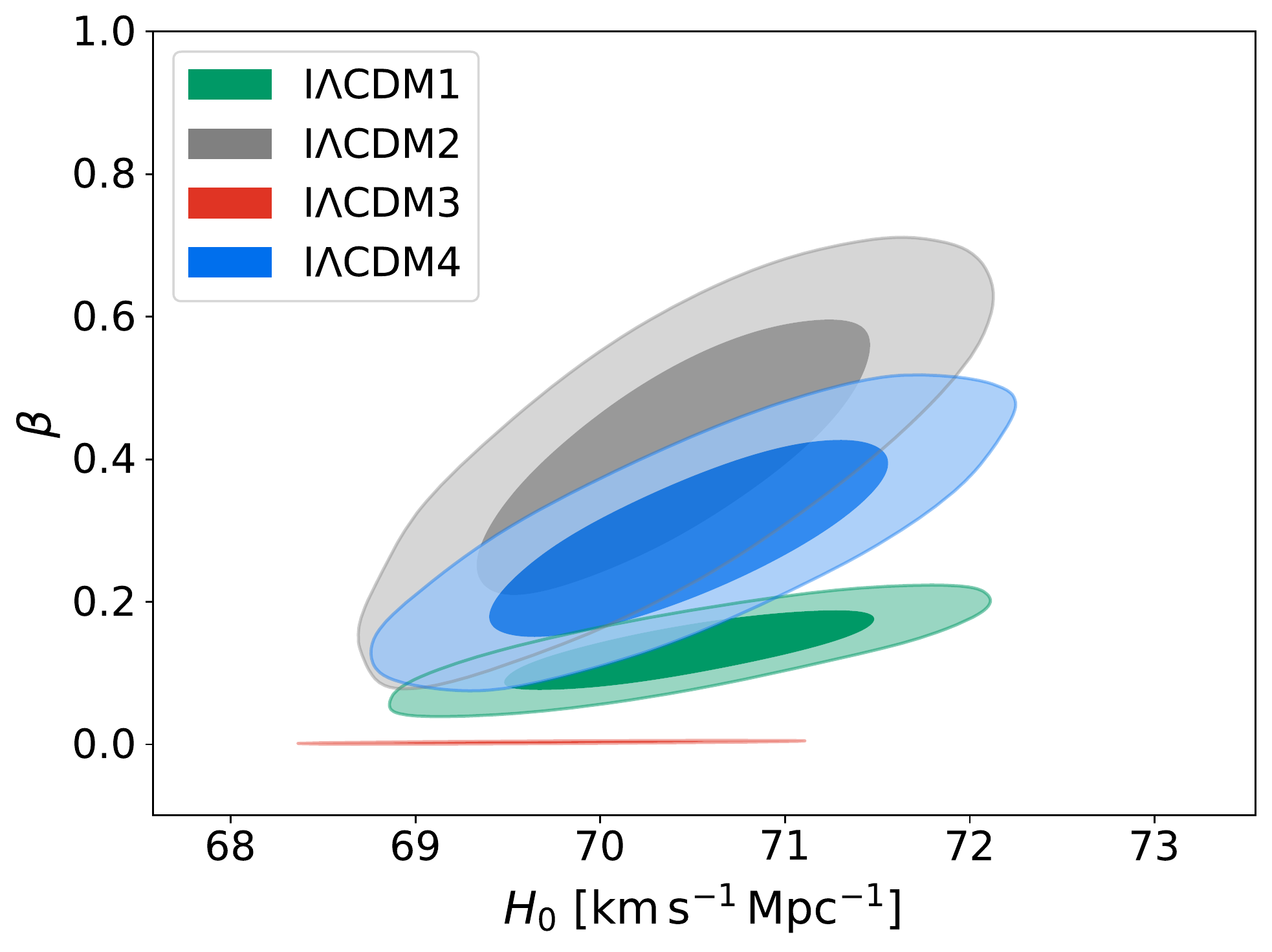}
    \caption{2D marginalized contours (68.3 and 95.4 per cent confidence levels) in the $H_{0}$--$\beta$ plane for the IDE models by using the CBSH$+$TD data.}
    \label{fig2}
\end{figure}

As for the coupling parameter $\beta$, the TD data alone cannot constrain it well. The TD data are actually the combination of the angular diameter distance that is inversely proportional to $H_0$, and thus the TD data are more sensitive to $H_0$ than to $\beta$. Therefore, we have to combine the TD data with the current mainstream observations, i.e., the CMB, BAO, SN, and $H_0$ data, to constrain the coupling parameter $\beta$.
Although in principle there is an inconsistency between the TD data and the CMB data in constraining $H_0$, we still give the joint constraints for completeness of the analysis. Here, we consider the local $H_0$ measurement reported by the SH0ES team as a prior and add it to the CMB$+$BAO$+$SN data. It should be added that a more reasonable approach to combine the SN and $H_0$ data is to adopt the local prior of SN Ia absolute magnitude instead of the corresponding prior of $H_0$, which can avoid double counting of low-redshift supernovae \citep[e.g.][]{Camarena:2019moy,Camarena:2019rmj,Camarena:2021jlr}.

When using the CBSH data without the TD data, the constraint values of the $\beta$ values are $0.095\pm 0.040$ (I$\Lambda$CDM1), $0.33\pm 0.14$ (I$\Lambda$CDM2), $0.0021\pm 0.0011$ (I$\Lambda$CDM3), and $0.225\pm 0.093$ (I$\Lambda$CDM4). In all the four IDE models, the $\beta$ values are positive, corresponding to the case of cold dark matter decaying into dark energy. In I$\Lambda$CDM3, the value of $\beta$ is close to zero, meaning that the CBSH data support no interaction between dark energy and cold dark matter in this model. When using the CBSH$+$TD data, the $\beta$ values are $0.132\pm 0.037$ (I$\Lambda$CDM1), $0.41\pm 0.13$ (I$\Lambda$CDM2), $0.0031\pm 0.0011$ (I$\Lambda$CDM3), and $0.293\pm 0.092$ (I$\Lambda$CDM4). It is shown that, when the TD data are added, the errors of $\beta$ are only slightly reduced, because the TD data cannot constrain $\beta$ tightly. The central values of $\beta$ become higher, because $\beta$ and $H_0$ are positively correlated and adding the TD data makes the $H_0$ values higher, which can be seen from Fig.~\ref{fig2}.
Higher positive values of $\beta$ further support the scenario that cold dark matter decays into dark energy.

Now, we give a discussion on the parameter $S_8$, defined as $S_8\equiv \sigma_8\sqrt{\Omega_{\rm m}/0.3}$, with $\sigma_8$ being the amplitude of mass fluctuations. 
The GC and WL data from DES give the results of $S_8=0.773^{+0.026}_{-0.020}$ (DES Year 1) \citep{DES:2017myr} and $S_8=0.776^{+0.017}_{-0.017}$ (DES Year 3) \citep{DES:2021wwk}, which are in more than 2$\sigma$ tension with $S_8=0.834^{+0.016}_{-0.016}$ inferred from the $\emph{Planck}$ CMB data \citep{Aghanim:2018eyx}.
{In previous works \citep[e.g.][]{DiValentino:2019ffd,DiValentino:2019jae,DiValentino:2020vvd,Gao:2021xnk,Lucca:2021dxo,Lucca:2021eqy,Abdalla:2022yfr,deAraujo:2021cnd}, the IDE models are considered to relieve the $S_8$ tension.} We also wish to investigate the effect of the TD data on $S_8$ in this work. However, the TD data alone cannot provide direct constraints on the parameter $S_8$, due to the fact that the time-delay distance provides only the geometrical information of the Universe, while constraining $S_8$ needs the information of the large-scale structure. Even so, when combined with other observational data, the TD data may indirectly affect the constraints on $S_8$ through constraining other cosmological parameters. Therefore, we discuss the effect of the TD data on $S_8$ in the context of the CBSH and CBSH$+$TD data. Since the $S_8$ values inferred from the DES Year 1 and DES Year 3 data have no obvious difference; in this paper, we employ only the DES Year 1 data for the cosmological analysis and leave the DES Year 3 data for future works. 


\begin{table*}
    \renewcommand\arraystretch{1.5}
	\centering
	\caption{Fitting results (68.3 per cent confidence level) in the $\Lambda$CDM and IDE models from the CBSH and CBSH$+$TD data. Here, $H_0$ is in units of ${\rm km}~{\rm s}^{-1}~{\rm Mpc}^{-1}$.}
	\label{CBSH}
	\begin{tabular}{lcccccc}
	    \hline
	    Data & Paratemer & $\Lambda$CDM & I$\Lambda$CDM1 & I$\Lambda$CDM2 & I$\Lambda$CDM3 & I$\Lambda$CDM4\\
	    \hline
	    CBSH&$H_{0}$&$68.20\pm 0.42$ & $69.57\pm 0.73$ & $69.58\pm 0.75$ & $68.96\pm 0.60$ & $69.65\pm 0.74$\\
	    &$\Omega _{m}$&$0.3034\pm 0.0055$ & $0.274\pm 0.013$ & $0.230\pm 0.032$ & $0.2950\pm 0.0071$ & $0.241\pm 0.026$\\
	    &$\beta$&$-$ & $0.095\pm 0.040$ & $0.33\pm 0.14$ & $0.0021\pm 0.0011$ & $0.225\pm 0.093$\\
	    &$\sigma_{8}$&$0.8049\pm 0.0072$ & $0.845\pm 0.018$ & $1.047^{+0.088}_{-0.160}$ & $0.820\pm 0.011$ & $0.990^{+0.072}_{-0.110}$\\
	    &$S _{8}$&$0.809\pm 0.012$ & $0.807\pm 0.012$ & $0.906^{+0.038}_{-0.060}$ & $0.813\pm 0.012$ & $0.882^{+0.031}_{-0.042}$\\
	    &$S_{8}$ tension&$1.59\sigma$ & $1.38\sigma$ & $1.63\sigma$ & $1.47\sigma$ & $1.34\sigma$\\
	    &$\chi_{\rm min}^2$&$3825.487$ & $3822.547$ & $3825.248$ & $3825.206$ & $3825.253$\\
	    &$\triangle$ AIC& $0$ & $-0.940$ & $1.760$ & $1.719$ & $1.765$\\
	    \hline
	    CBSH$+$TD&$H_{0}$&$68.67\pm 0.41$ & $70.48\pm 0.67$ & $70.40\pm 0.71$ & $69.72\pm 0.56$ & $70.50\pm 0.71$\\
	    &$\Omega _{m}$&$0.2975\pm 0.0053$ & $0.259\pm 0.011$ & $0.205\pm 0.031 $ & $0.2862\pm 0.0064$ & $0.217\pm 0.026$\\
	    &$\beta$&$-$ & $0.132\pm 0.037$ & $ 0.41\pm 0.13$ & $0.0031\pm 0.0011 $ & $0.293\pm 0.092$\\
	    &$\sigma_{8}$&$0.8026\pm 0.0073$ & $0.858\pm 0.018$ & $ 1.14^{+0.10}_{-0.19}$ & $ 0.826\pm 0.011 $ & $1.072^{+0.079}_{-0.140} $\\
	    &$S _{8}$&$0.799\pm 0.012$ & $0.798\pm 0.012$ & $0.932^{+0.043}_{-0.069}$ & $0.807\pm 0.012$ & $0.904^{+0.034}_{-0.050}$\\
	    &$S_{8}$ tension&$1.16\sigma$ & $1.11\sigma$ & $1.86\sigma$ & $1.26\sigma$ & $1.58\sigma$\\
	    &$\chi_{\rm min}^2$&$3844.530$ & $3835.107$ & $3843.417$ & $3840.304$ & $3844.153$\\
	    &$\triangle$ AIC& $0$ & $-7.423$ & $0.887$ & $-2.226$ & $1.622$\\
        \hline
	\end{tabular}
\end{table*}

\begin{figure*}
    \centering
    \subfloat[]{\label{fig4a}\includegraphics[width=0.65\columnwidth]{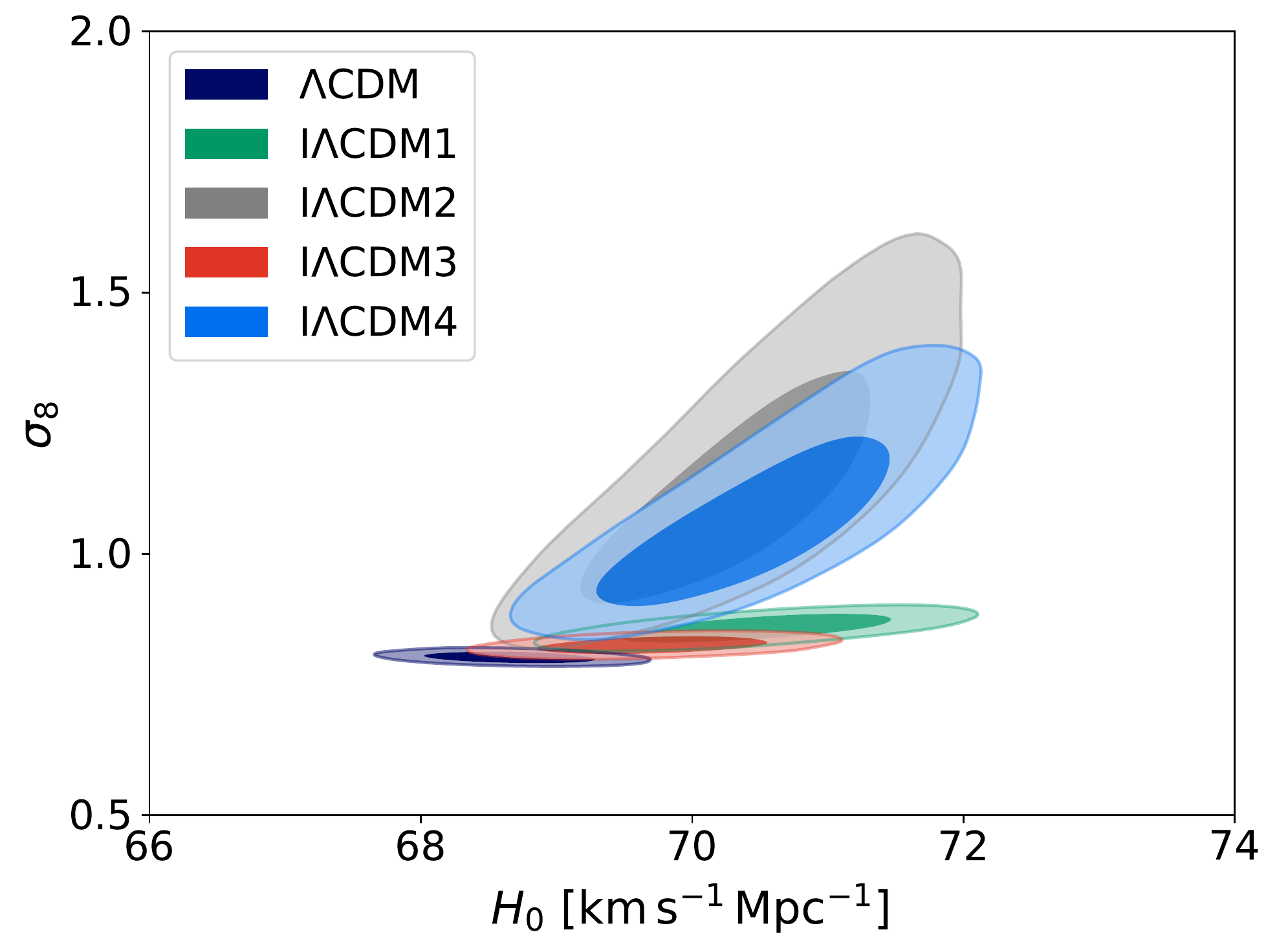}}
    \subfloat[]{\label{fig4b}\includegraphics[width=0.65\columnwidth]{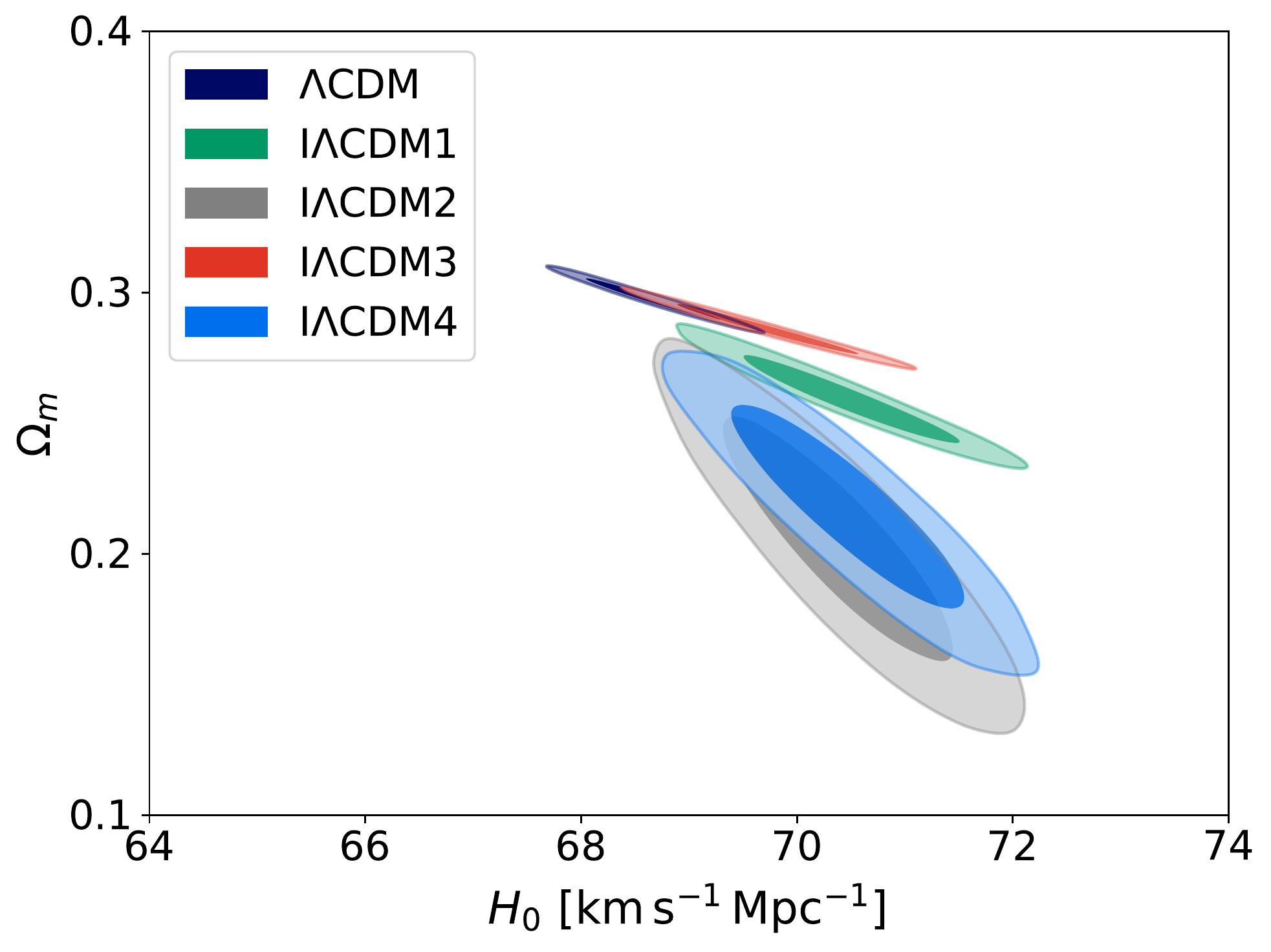}}
	\subfloat[]{\label{fig4c}\includegraphics[width=0.65\columnwidth]{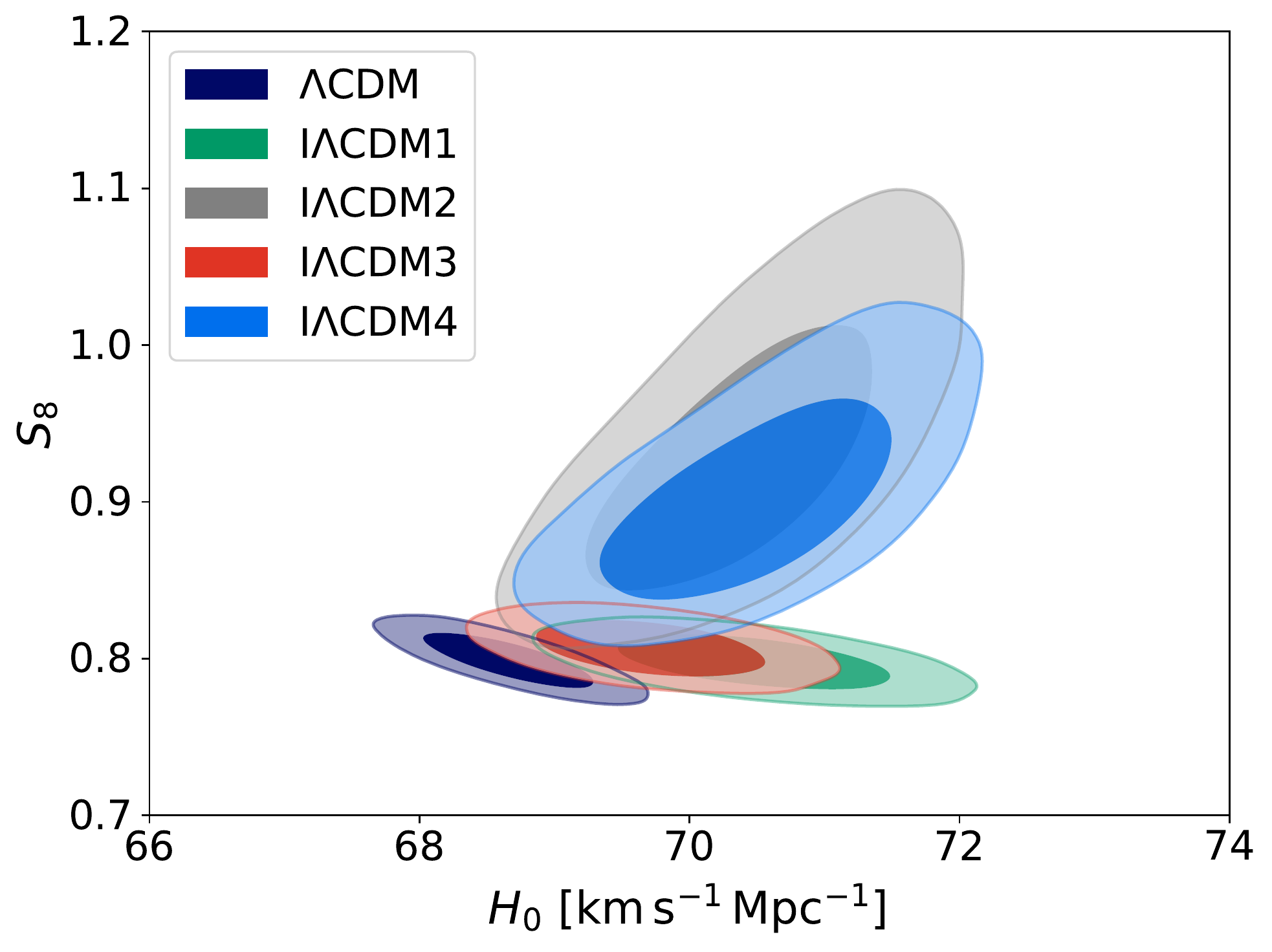}}
    \caption{2D marginalized contours (68.3 and 95.4 per cent confidence levels) in the $H_{0}$--$\sigma_{8}$, $H_{0}$--$\Omega_{\rm m}$, and $H_{0}$--$S_{8}$ planes for the $\Lambda$CDM model and the IDE models by using the CBSH$+$TD data.} 
    \label{fig4}
\end{figure*}

When the GC and WL data are employed, the
inferred $S_8$ values are $0.772^{+0.019}_{-0.021}$ ($\Lambda$CDM), $0.760^{+0.034}_{-0.030}$ (I$\Lambda$CDM1), $0.785^{+0.048}_{-0.064}$ (I$\Lambda$CDM2), $0.771\pm 0.026$ (I$\Lambda$CDM3), and $0.786^{+0.054}_{-0.069}$ (I$\Lambda$CDM4). We calculate the $S_8$ tensions by comparing these inferred $S_8$ values with the CBSH and CBSH$+$TD results reported in Table~\ref{CBSH}.
When the CBSH data are employed, the $S_8$ tensions in the I$\Lambda$CDM1, I$\Lambda$CDM3, and I$\Lambda$CDM4 models are 1.38$\sigma$, 1.47$\sigma$, and 1.34$\sigma$, respectively, slightly relieved compared with 1.59$\sigma$ in the $\Lambda$CDM model. 
When the CBSH$+$TD data are employed, compared with the $\Lambda$CDM model, almost all the considered IDE models further aggravate the $S_8$ tension, except the I$\Lambda$CDM1 model very slightly relieving the tension from 1.16$\sigma$ to 1.11$\sigma$. The slight alleviation of the $S_8$ tension is due to the fact that $S_8$ is determined by both $\sigma_8$ and $\Omega_{\rm m}$. Although higher $H_0$ leads to higher $\sigma_8$, $\Omega_{\rm m}$ becomes lower at the same time, as shown in Fig.~\ref{fig4}\subref{fig4a} and \ref{fig4}\subref{fig4b}. These two opposite effects eventually make $S_8$ become lower, as shown in Fig.~\ref{fig4}\subref{fig4c}.

Finally, we compare the IDE models on the basis of their fittings to the observational data. It is unfair to use only $\chi^{2}_{\rm min}$ to judge how well the models can fit the data, because the IDE models have one more free parameter ($\beta$) than the $\Lambda$CDM model. Hence, we use the Akaike information criterion \citep[AIC;][]{Szydlowski:2008by}, defined as $\rm{AIC}\equiv\chi^2_{\rm min}$+2$d$, with $d$ being the number of free parameters, to compare the fittings. In order to show the differences of AIC values between the $\Lambda$CDM and IDE models more clearly, we set the AIC value of the $\Lambda$CDM model to be zero, and list the values of $\Delta {\rm AIC}=\Delta \chi^2+2\Delta d$ in Table~\ref{CBSH}, with $\Delta \chi^2=\chi^{2}_{\rm min, IDE}-\chi^{2}_{\rm min, \Lambda CDM}$ and $\Delta d=d_{\rm IDE}-d_{\rm \Lambda CDM}=1$. A model with a lower value of AIC is more supported by the observational data.

When the CBSH data are employed to constrain the IDE models, the $\Delta$AIC values of the IDE models are $-0.940$ (I$\Lambda$CDM1), 1.760 (I$\Lambda$CDM2), 1.719 (I$\Lambda$CDM3), and 1.765 (I$\Lambda$CDM4). Only the I$\Lambda$CDM1 model has a slightly lower AIC than the $\Lambda$CDM model. The absolute values of $\Delta$AIC in all the four IDE models are not large, indicating that the CBSH data support the IDE and $\Lambda$CDM models to the similar extent. However, this situation changes significantly when the TD data are added. When we use the CBSH$+$TD data, the $\Delta$AIC values of the I$\Lambda$CDM1 and I$\Lambda$CDM3 models dramatically decrease, with $\Delta {\rm AIC}=-7.423$ (I$\Lambda$CDM1) and $\Delta {\rm AIC}=-2.226$ (I$\Lambda$CDM3). This indicates that the data sets including the TD data support the IDE models with $Q\propto\rho_{\rm c}$ more than those with $Q\propto\rho_{\rm de}$. Especially for the I$\Lambda$CDM1 model ($Q=\beta H_0 \rho_{\rm c}$), its AIC value is significantly lower than that of the $\Lambda$CDM model. We can conclude that this model has more advantages than the $\Lambda$CDM model, in terms of fitting the observational data.

\section{Conclusions}\label{sec:con}
Time-delay cosmography provides an important complement to the late-Universe observations, based on the time-delay effect of strong gravitational lensing, coding the information of the time-delay distance $D_{\Delta t}$ defined as a combination of angular diameter distances. In this work, we investigate the implications of the time-delay cosmography on the IDE models. The measurements of time-delay cosmography from seven lensed quasars, abbreviated as the TD data, are considered. Four representative IDE models, i.e., the I$\Lambda$CDM1 ($Q = \beta H_{0} \rho_{\rm c}$), I$\Lambda$CDM2 ($Q = \beta H_{0} \rho_{\rm de}$), I$\Lambda$CDM3 ($Q = \beta H \rho_{\rm c}$), and I$\Lambda$CDM4 ($Q = \beta H \rho_{\rm de}$) models, are considered in this work. We first employ the TD data alone to constrain the IDE models, and then we combine the TD data with the local $H_0$ measurement by the SH0ES team to address the $H_0$ tension. Then we combine the TD data with the CMB$+$BAO$+$SN$+H_0$ data to give constraints on the coupling parameter $\beta$, and discuss the $S_8$ tension. Finally, we discuss the comparison of the IDE models according to the fitting results. The main findings from our analyses are summarized as follows.

\begin{itemize}

\item[(i)] When the TD data alone are employed, the $H_0$ tensions between the CMB and TD data are reduced from 3.88$\sigma$ ($\Lambda$CDM) to $1.91\sigma$ (I$\Lambda$CDM2) and $1.64 \sigma$ (I$\Lambda$CDM4), respectively. When combining the TD data and the local $H_0$ measurement by the SH0ES team, the $H_0$ tensions are reduced from 5.37$\sigma$ ($\Lambda$CDM) to $2.06\sigma$ (I$\Lambda$CDM2) and 1.67$\sigma$ (I$\Lambda$CDM4). This implies that the IDE models with the interaction term $Q \propto \rho_{\rm de}$ seem to have an advantage in relieving the $H_0$ tension between the CMB and TD data. However, it is hard to draw a firm conclusion, because the reductions of the $H_0$ tensions are mainly due to the relatively large errors of $H_0$ inferred from the CMB data in the IDE models.

\item[(ii)] The TD data alone cannot constrain the coupling parameter $\beta$ well. Adding the TD data to the CBSH data, the central values of $\beta$ increase due to the positive correlation between the parameters $\beta$ and $H_0$. The $\beta$ values are $0.132\pm 0.037$ (I$\Lambda$CDM1), $0.41\pm 0.13$ (I$\Lambda$CDM2), $0.0031\pm 0.0011$ (I$\Lambda$CDM3), and $0.293\pm 0.092$ (I$\Lambda$CDM4) when the CBSH$+$TD data are employed. The $\beta$ values are positive within 1$\sigma$ range in the four IDE models, corresponding to the case of cold dark matter decaying into dark energy.

\item[(iii)] When the CBSH$+$TD data are employed, only the I$\Lambda$CDM1 model can relieve the $S_8$ tension very slightly, while the other IDE models further aggravate the $S_8$ tension. This indicates that the IDE models have no obvious advantage in relieving the $S_8$ tension, although the IDE model with $Q = \beta H_0 \rho_{\rm c}$ shows a mild advantage.

\item[(iv)] When the CBSH data are employed, the $\Delta$AIC values of the IDE models are not very different, but when the TD data are added, the $\Delta$AIC values of the I$\Lambda$CDM1 and I$\Lambda$CDM3 model are obviously lower than the other IDE models, especially for the I$\Lambda$CDM1 model ($Q=\beta H_0 \rho_{\rm c}$) with $\Delta {\rm AIC}=-7.423$. From the perspective of fitting the observational data, the IDE model with $Q=\beta H_0 \rho_{\rm c}$ is more supported by the CBSH$+$TD data than the $\Lambda$CDM model.

\end{itemize}

In summary, the IDE models with $Q \propto \rho_{\rm de}$ seem to have an advantage in relieving the $H_0$ tension between the early-Universe and late-Universe measurements; all the considered IDE models have no obvious advantage in relieving the $S_8$ tension; the IDE models with $Q \propto \rho_{\rm c}$ have advantages in fitting the observational data. None of the considered IDE models can perform well in all aspects. The performance of more potential IDE models in the TD data will be studied in our future work. In the future, more TD data will be obtained, helping us use late-Universe cosmological probes to further study the interaction between dark energy and dark matter.


\section*{Acknowledgements}
This work was supported by the National Natural Science Foundation of China (grant numbers 11975072, 11835009, 11875102, and 11690021), the Liaoning Revitalization Talents Program (grant number XLYC1905011), the Fundamental Research Funds for the Central Universities (grant number N2005030), the National Program for Support of Top-Notch Young Professionals (grant number W02070050), the National 111 Project of China (grant number B16009), and the Science Research Grants from the China Manned Space Project (grant number CMS-CSST-2021-B01).

\section*{Data Availability}
The data underlying this article will be shared on reasonable request
to the corresponding author.



\bibliographystyle{mnras}
\bibliography{idelens} 








\bsp	
\label{lastpage}
\end{document}